\documentclass[11pt]{article}

\usepackage[left=2.5cm, right=2.5cm, top=2.5cm, bottom=2.5cm]{geometry}
\usepackage{url}

\usepackage{listings}
\usepackage{multirow}

\usepackage{xcolor}

\usepackage[hidelinks, colorlinks=true, linkcolor=blue, citecolor=blue, urlcolor=blue]{hyperref}

\newcommand{\probone}{P1}
\newcommand{\probtwo}{P2}
\newcommand{\probthree}{P3}
\newcommand{\probfour}{P4}
\newcommand{\probfive}{P5}
\newcommand{\userqone}{Q1}
\newcommand{\userqtwo}{Q2}
\newcommand{\userqthree}{Q3}
\newcommand{\gptthree}{GPT-3.5T}
\newcommand{\gptfour}{GPT-4T}

\newcommand{\citep}[1]{\cite{#1}}

\begin{document}


\title{Prompt-Based Cost-Effective Evaluation and Operation of ChatGPT as a Computer Programming Teaching Assistant}

\author{Marc Ballestero-Ribó\qquad Daniel Ortiz Martínez\\{\small Department of Mathematics and Computer Science, Universitat de Barcelona, Barcelona, Spain}}

\date{}

\maketitle

\begin{abstract}
The dream of achieving a student-teacher ratio of 1:1 is closer than ever thanks to the emergence of large language models (LLMs). One potential application of these models in the educational field would be to provide feedback to students in university introductory programming courses, so that a student struggling to solve a basic implementation problem could seek help from an LLM available 24/7. This article focuses on studying three aspects related to such an application.  First, the performance of two well-known models, \gptthree\ and \gptfour, in providing feedback to students is evaluated. The empirical results showed that \gptfour\ performs much better than \gptthree, however, it is not yet ready for use in a real-world scenario. This is due to the possibility of generating incorrect information that potential users may not always be able to detect.  Second, the article proposes a carefully designed prompt using in-context learning techniques that allows automating important parts of the evaluation process, as well as providing a lower bound for the fraction of feedbacks containing incorrect information, saving time and effort. This was possible because the resulting feedback has a programmatically analyzable structure that incorporates diagnostic information about the LLM's performance in solving the requested task. Third, the article also suggests a possible strategy for implementing a practical learning tool based on LLMs, which is rooted on the proposed prompting techniques. This strategy opens up a whole range of interesting possibilities from a pedagogical perspective.
\end{abstract}

\section{Introduction}\label{sec:intro}
Achieving a 1:1 student-to-teacher ratio holds the promise of deeply alter educational environments. With this personalized approach, teaching methods could be tailored to each student, accommodating the diverse learning styles, abilities, and preferences that exist within a classroom. It would also facilitate timely feedback on assignments, enabling students to learn from their mistakes and make continuous improvements.  Unfortunately, having a 1:1 student-to-teacher ratio has been an elusive goal due to the significant resource challenges that would be inherently involved, from the necessity of finding qualified teachers or additional classroom spaces to the tremendous financial impact that implementing this model would have on the limited budgets of educational institutions.

One good example of this problematic are the programming courses during the first year of Computer Engineering degrees. Such courses have a crucial role in the basic training of students as future professionals, since for many of them, it is within those courses where they are in contact with the task of programming for the first time. However, the learning process for students within these initial courses is not straightforward. In the first year of a regular Computer Engineering degree, the number of enrolled students can be expected to be very high (e.g., more than 100 students enrolled in the Computer Engineering degree of the Universitat de Barcelona during the 2023-2024 academic year). Furthermore, in this context characterized by a high concentration of students, it would be ideal for each one to be able to solve a large number of basic code implementation exercises (e.g., define a Python function to calculate the factorial of a number given as input), always with the assistance and feedback provided by the course's faculty. This assistance should be available, not only during teaching sessions or office hours, but also when the students work independently. During that time, if the students encounter difficulties in solving coding problems, they should contact the teaching staff via email. This process may be time-consuming and make the learning less effective.

The ideal situation where each student receives constant and personalized assistance is far from being achievable, given that the number of teachers who are available to attend the students' needs is limited. However, the emergence of artificial intelligence (AI) technologies based on large language models (LLMs) makes it possible to aspire to a new scenario in which learning tools could be offered to students, always within a restricted and controlled environment, acting as virtual teaching assistants for the course and thereby bringing closer the goal of a 1:1 student-to-teacher ratio. These teaching assistants would be available 24/7 to address programming queries from students. The assistance provided by LLMs would aim to guide the student to the solution of programming problems, rather than simply providing the solution itself.

There already exist prominent examples of LLMs being used as coding assistants. The most popular of them being GitHub Copilot~\cite{githubcopilot21}. GitHub Copilot is an innovative AI-powered coding assistant developed by GitHub in collaboration with OpenAI. It serves as an extension to integrated development environments (IDEs) and assists developers in writing code by providing context-aware code suggestions and completions. However, it is important to stress out that GitHub Copilot would not be aligned with the specific necessities that a virtual teaching assistant of initial undergraduate computer programming courses would need to cover. As it was explained above, the ideal LLM-based assistant that would be required should be focused on guiding the student to the solution instead of providing it. This contrasts with the general purpose of tools like GitHub Copilot, which are more targeted to individuals who are already able to write code and want to increase their productivity.

The most prominent LLM technology available so far that has a strong potential to be applied as a teaching assistant is ChatGPT~\cite{brown20lma}. ChatGPT is a chatbot built on top of an LLM technology called generative pre-training transformer~\cite{radford18ilu} (GPT). ChatGPT is able to understand and generate human-like text based on the input it receives. ChatGPT has demonstrated a remarkable ability in different tasks, such as answering questions, providing information or engaging in dynamic conversations.

However, using ChatGPT as a teaching assistant presents several challenges. One of them would be that ChatGPT is a general purpose model. In spite of the fact that the model's training data encompasses a vast array of internet text, it cannot be expected to have the course-specific depth and accuracy required for a dedicated teaching assistant role. Its general purpose nature may also hinder its ability to adapt its responses to diverse learning styles or to provide pedagogically sound explanations, where it is of utmost importance that the model stimulates the academic growth of the students instead of merely handing requested information to them. Another challenge to apply ChatGPT as a teaching assistant is the necessity to know what this technology can and cannot do in its present form when applied for that purpose. Although there are countless examples where ChatGPT as well as other LLMs have already demonstrated a remarkable ability to solve complex tasks, their outputs are not immune to errors, biases, or misunderstandings.

Evaluating the performance of ChatGPT's or other LLMs is a challenging task. Ideally, evaluation should be carried out by human experts, generating one or multiple manual evaluation metrics that capture aspects related to the performance or quality of the model's responses. However, manual evaluation metrics present significant drawbacks, with the main one being their high cost in resources. This problem is even more serious in the extremely dynamic ecosystem of LLMs today, where new models or variants of them requiring evaluation are constantly being made available to the public.

One possible way to alleviate this problem is through the acquisition of automatic evaluation measures. Automatic measures are interesting because they are cheap and quick to obtain, reproducible, and typically provide quantitative results that allow for comparing different models. However, automatic measures may fail to give an accurate idea of the model's performance if they are not specifically designed for a particular task. One good example of this would be the different LLMs leaderboards created so far reporting measures that may be only vaguely related with a particular LLM application of interest.

Therefore, it would be desirable to find automatic measures specifically designed to evaluate the performance of LLMs as teaching assistants in undergraduate computer programming courses, studying the relationship between such measures and those ones that cover other aspects that can only be evaluated manually.

\begin{sloppypar}
In spite of the fact that ChatGPT has already been evaluated in an academic context~\citep{laskar23ass,jacobsen23tpa,savelka23cgp}, only a few studies are focused on evaluating the performance of the tool when working with code~\citep{hellas23etr,macneil23dle,jalil23cas}. However, to the best of our knowledge, none of them address the problem of the evaluation cost.
\end{sloppypar}

The goal of this work is threefold. First, it focuses on evaluating whether ChatGPT is ready to be used as a teaching assistant within undergraduate computer programming courses. Second, it defines a methodology to analyze the feedback provided by ChatGPT in a programmatic manner, producing automatic evaluation measures whose relation with manual ones can be analyzed. Finally, the third goal of the paper goes beyond evaluation, sketching a possible way to effectively operate ChatGPT as a computer programming teaching assistant in a real scenario. This proposal is strongly intertwined with the above-mentioned methodology to analyze ChatGPT's feedback.

\section{Research Questions}\label{sec:res_ques}
In order to investigate the ability of ChatGPT to function as a teaching assistant in undergraduate computer programming courses, we will explore the following research questions:

\paragraph{RQ1} Is ChatGPT able to determine whether a student submission is correct or not according to the problem definition? Being able to understand the student code is a prerequisite for the subsequent explanation of the potential code errors.

\paragraph{RQ2} Is ChatGPT able to propose a new version of a faulty student submission that is correct according to the problem definition? This is strongly related to the previous point, since, in order to guide a student towards obtaining correct code, another prerequisite would be the ability to generate a correct solution starting from the defective code.

\paragraph{RQ3} Is ChatGPT able to successfully identify the errors present in a student submission? This constitutes the main task that an application implementing an LLM-based teaching assistant should carry out.

\paragraph{RQ4} Are the students able to discern if ChatGPT's feedback about a particular exercise is correct or not? A well-known problem with ChatGPT is the possibility of generating responses that appear to be correct but are not actually accurate. This issue is particularly relevant when considering the use of ChatGPT as a teaching assistant, as it may promote the assimilation of incorrect ideas by the student, whose ability to determine the validity of the information provided is still in development.

\paragraph{RQ5} Is the structure of ChatGPT's responses consistent enough to be analyzed in a fully automated way? The answer to this question has important implications regarding the ability to generate automatic evaluation measures and also when operating the model in real scenarios, as will be explained in Section~\ref{subsec:eff_oper}.

\paragraph{RQ6} Provided that it is possible to automate parts of the evaluation process, are there relationships between automatic and manual evaluation measures? Establishing those relationships would be useful to enable simple and inexpensive measurement of the performance of a specific version of ChatGPT and compare it with other ones, or even with other LLM families.

\section{Methodology}\label{sec:methodology}

\subsection{Tested Models}\label{subsec:tested_models}
This work focused on comparing the performance of two OpenAI's LLMs\footnote{\url{https://platform.openai.com/docs/models/overview}} for their use as teaching assistants:
\begin{itemize}
    \item {\bf GPT-3.5 Turbo (\gptthree)}: a family of LLMs that take textual information as input and generate text as output. According to the OpenAI documentation, they constitute improved versions of GPT-3.5.
    \item {\bf GPT-4 Turbo (\gptfour)}: they are large scale, multimodal models that accept image and text inputs, and produce text outputs~\citep{openai23gpt4}. These models are also described as improved versions of GPT-3.5 in the OpenAI's documentation.
\end{itemize}

In all cases, the models were operated by means of the application programming interface (API) provided by OpenAI.

\subsection{Data Acquisition}\label{subsec:data_acquisition}
The experiments presented in this article are based on the solutions to five Python programming problems provided by first-year computer science students at the University of Barcelona. These problems were presented to the students in an evaluative exam during the academic years 2022/2023 and 2023/2024 within the subject of algorithms.

More specifically, the two problems to be solved were:
\begin{itemize}
    \item {\bf Rotated Palindromes (\probone)}: A Python function should be implemented that receives a string as input. The function should return \lstinline{True} if there is any rotation of the input that constitutes a palindrome and \lstinline{False} otherwise. Given a string, a rotation is defined as a new string where the last character is removed and inserted again at the start. The function should return \lstinline{True} if at least one of the rotations is a palindrome. For instance, if the input string is \lstinline{AAB}, the function should return \lstinline{True}, since one of the rotations (\lstinline{ABA}) is a palindrome. On the other hand, for the input string \lstinline{AB}, the function should return \lstinline{False}, since there are no palindromic rotations.
    \item {\bf Run Length Encoding (\probtwo)}: Again, a Python function receiving a string as input should be implemented. The function should return the run-length-encoding (RLE) compression of the input string. In particular, the function should return a new string that for each character incorporates the number of consecutive repetitions. For instance, if the input string is \lstinline{AAABB}, then the function should return \lstinline{A3B2}.
    \item {\bf Number of Ones (\probthree)}: Given an ordered list consisting of only the numbers 0 and 1, a Python function should be implemented that calculates the number of ones contained in the list. The function should take advantage of the fact that the list is sorted to improve efficiency; otherwise, it will be considered erroneous.
    \item {\bf In-place Partition (\probfour)}:  Given a list $A$ and a value $v$, a Python function should be implemented that changes the order of the elements so that at the beginning all elements equal to $v$ appear, then all elements smaller than $v$, and finally all elements greater than $v$. The result should be generated in-place in the input list $A$. The elements in each of the three parts that compose the list may appear in any order.
    \item {\bf Sum of Pairs (\probfive)}:  A Python function should be implemented that, given an ordered list of non-repeating integers $L$ and a value $s$, finds all pairs of numbers in the list that add up to this value.
\end{itemize}

The students were given Python notebooks containing these five problems as well as others within the evaluative exams, and were asked to submit the notebook with the solutions through an application at the end of the exam.

The chosen scenario allows simulating a learning context like the one described in Section~\ref{sec:intro}, in which a group of students faces a specific programming problem and requests help on it. In particular, we assume that the student has asked for feedback on their implementation exactly in the state in which the implementation was at the time of exercise submission. This ensures that the evaluation can be done for very diverse implementations, ranging from those that work correctly to those that cannot even be executed due to compilation or runtime errors.

\subsection{User Evaluation}\label{subsec:user_eval}
In this study, we specifically acquired data involving a small group of potential end users of the LLM as a teaching assistant, in connection with RQ4. For this purpose, two implementations for the above-described problems \probone\ and \probtwo\ (cases \emph{a} and \emph{b}) were analyzed with the \gptfour\ model, providing feedback about the code issues. From the four implementations, \gptfour\ pointed out really existent issues with the code in two cases (one for problem \probone\ and one for problem \probtwo), and mentioned non-existent issues for the other two. Below there is a description of the main features of each case:
\begin{itemize}
\item {\bf \probone\ case \emph{a}}: the student implementation was wrong and ChatGPT provided correct feedback about the errors.
\item {\bf \probone\ case \emph{b}}: the student presented a faulty implementation, but ChatGPT's feedback did not correctly identify the errors.
\item {\bf \probtwo\ case \emph{a}}: the student's solution was incorrect and ChatGPT successfully identified some (but not all) errors.
\item {\bf \probtwo\ case \emph{b}}: the student solution was correct but ChatGPT pointed out non-existing problems in its feedback.
\end{itemize}

The experimentation was restricted to \probone\ and \probtwo because these two problems in combination with cases \emph{a} and \emph{b} provided a complete range of situations arising when students work with code.

The four implementations with their corresponding feedback about code issues were given to a subgroup of students belonging to the above-mentioned first-year computer science course for evaluation. The subgroup was composed of 11 students (8 men and 3 women). Prior to this, a face-to-face meeting to inform about the study was organized. After the meeting, those students willing to participate were sent the above-mentioned 4 Python notebooks for their asynchronous evaluation, with the intention to create an environment similar to that where a real LLM-based tool would be used (that is, during the students' autonomous work). The students were explicitly told that the code to be analyzed could contain errors or not. For each notebook, the students responded to three questions with no time limitation:

\begin{itemize}
\item {\bf Question 1 (\userqone)}: \emph{Do you think that the feedback provided by ChatGPT is correct, that is, does it identify real errors present in the code?}
\item {\bf Question 2 (\userqtwo)}: \emph{If you answered yes to Question 1, to what extent is the feedback provided by ChatGPT useful to you to correct the errors present in the code? Respond with a number from 1 to 5, where 1 means that you found the help very little useful and 5 means that the help was very useful.}
\item {\bf Question 3 (\userqthree)}: \emph{Could you explain what led you to decide whether ChatGPT's feedback was correct or not? Examples: ``I modified the code based on the feedback and ran it myself'', ``the words used by ChatGPT gave me confidence'', ``I noticed that ChatGPT said the code was doing something it wasn't doing'', ``ChatGPT help was difficult to understand'', etc.}
\end{itemize}

Finally, a total of 40 questionnaires were collected, since not all of the students sent back their responses to the four notebooks. In summary, for \probone, 11 notebooks were collected for case \emph{a} and 9 for case \emph{b}. Regarding \probtwo, 10 notebooks were collected for case \emph{a} and another 10 for case \emph{b}.

\subsection{Characterizing Good LLM Feedback}\label{subsec_char_good_llm_feedback}

When using an LLM to provide feedback on programming problems, it is important to establish the qualities that would be desirable in such feedback under a pedagogic point of view. In this article, we identify the following:

\begin{itemize}
\item The feedback should guide towards the solution instead of providing it directly, which is key to driving the learning process. Therefore, good feedback should not contain code that implements the solution to a problem, but rather provide clues to resolve possible errors. Previous evaluations of LLMs in contexts similar to ours have shown that the GPT model family has a strong tendency to incorporate code that solves a problem even if the prompt used does not request it. For example, in~\citep{hellas23etr}, it is reported that the evaluated feedback almost always contained code, often related to the solution.
\item When evaluating the usefulness of LLM feedback, it is not as important from our perspective if it points out all the existing problems in a particular implementation. It is sufficient if it indicates a subset of them. At least in the personal experience of the author of this work, the interaction between the student and the teacher typically consists in the student iterating on his/her implementation with the help of the teacher, gradually approaching a correct solution.
\item It is particularly problematic for the student's learning if the feedback points out non-existent problems in the code being analyzed by the LLM. Since this possibility cannot be ruled out at all, as demonstrated in numerous evaluations of ChatGPT's responses in different areas, where the model sometimes just provide invented information, it would be of great interest to have mechanisms that could detect such situations in order to prevent the newly generated feedback from being shown to the student.
\end{itemize}

\subsection{Prompting Strategy}\label{subsec:prompting_strategy}
Prompt design constitutes a central aspect of this work. In the previous section, two important features of high-quality feedback from a pedagogical point of view were mentioned, whose treatment could be addressed by means of an appropriate prompt. These two aspects would be, on the one hand, avoiding the presence of code within the feedback, and on the other hand, introducing mechanisms that allow diagnosing problems with the feedback content, particularly if it points out the presence of non-existent errors in the code under analysis.

Both problems could benefit from an automatic treatment of the feedback content. In this way, if a response contained code, it could be removed provided that it could be effectively detected and extracted without altering the integrity of the rest of the feedback. On the other hand, problems with the reliability of feedback content might be detected early if it was possible to extract relevant information about the decisions made by the model to generate the feedback. Elements such as a description of the sequence of operations carried out by the analyzed code and whether the model considers that this sequence of operations solves the problem correctly or not could be helpful if they could be analyzed programmatically.

Previous works similar to ours have not emphasized the potential advantages of a carefully designed prompt to address the previously mentioned problems. Instead, in this article, the prompt design is particularly focused on producing feedback with a structure that is susceptible to be analyzed in a completely automated manner. To achieve this, the use of in-context-learning (ICL) techniques~\citep{dong23aso} is proposed, and more specifically, a variety of ICL called chain of thought (CoT)~\citep{wei23cot}.

The basic idea behind ICL is to include in the prompt itself a certain number of input-output pairs belonging to the task that one wants to solve through the LLM, aiming for the model to learn to perform its task by analogy (although ICL does not have the ability to directly influence the model's parameters). As a result, this strategy has been shown to improve the capability of LLMs in various tasks compared to the use of standard prompts. CoT further specializes the philosophy of ICL, so that the examples belonging to the task to be solved are no longer limited to providing input-output pairs but also the intermediate reasoning steps that lead to the originally sought answers. This approach has demonstrated significant improvements in tasks requiring complex reasoning, among which would be the analysis of code generated by students investigated in this work.

Figure~\ref{fig:prompt} shows the template used to construct the prompt adopted in this study. This template must be instantiated depending on the programming problem one wishes to analyze with the LLM. The prompt starts indicating the name of the function to be implemented as well as a brief description of it. The problem statement is closed with a list of unit tests or asserts that should be passed by a correct implementation. After that, an ICL approach is used, where a number of implementations with their corresponding feedbacks are provided as example. Only one sample implementation is shown in Figure~\ref{fig:prompt}, but the structure can be repeated as many times as desired. The feedback is structured in sections using Markdown language. The included sections are the following:

\begin{itemize}
\item {\bf Brief Code Explanation}: this section applies a CoT approach to decide whether a implementation is correct or not. For this purpose, it provides an enumeration describing the sequence of steps carried out by the implementation. After that, the verdict about correctness is given with a strictly defined format.
\item {\bf Main Issues}: if the implementation is judged as incorrect, the issues that are present are given as a list. Otherwise, the section will be void.
\item {\bf Corrected Version}: this section is also conditional to whether the implementation was deemed correct or not. The corrected version of the implementation is put here if the code was not correct; otherwise the section is left empty. In spite of the fact that in principle we are not interested in showing the solution to the student, this section is intended to provide a specific place to put it so that it can be easily filtered out without affecting the integrity of the rest of the elements of the LLM response. In addition to this, the analysis of the corrected version can provide useful information about the ability of the LLM to work with code.
\end{itemize}

\begin{figure}[h!]
\begin{lstlisting}[basicstyle=\small]
You are a teacher who should provide feedback for undergraduate computer programming
assignments. You will be provided with the code of a Python function implemented by a student
called <FUNCTION_NAME>. <FUNCTION_DESCRIPTION>.

The code should pass the following asserts:

<ASSERT_1>
<ASSERT_2>
...
<ASSERT_N>

Q: Please provide feedback for the following implementation of <FUNCTION_NAME>:

<SAMPLE_IMPLEMENTATION>

# Feedback

## Brief Code Explanation

1. <STEP_EXPLANATION_1>
2. <STEP_EXPLANATION_2>
...
N. <STEP_EXPLANATION_N>

Is the function correct according to the problem definition [YES/NO]? <RESPONSE>

## Main Issues (if the function is not correct)

- <ISSUE_1>
- <ISSUE_2>
- ...
- <ISSUE_N>

## Corrected Version (if the function is not correct)

<CORRECTED_CODE>

Q: Please provide feedback for the following implementation of <FUNCTION_NAME>:

<IMPLEMENTATION_TO_BE_ANALYZED>
\end{lstlisting}
  \caption{Prompt template for analysis of computer programming assignments.}
  \label{fig:prompt}
\end{figure}

There are two fundamental goals of the prompting strategy that is shown in Figure~\ref{fig:prompt}. First, to provide a fixed structure for the LLM's response, and second, to incorporate an explicit decision about the correctness of the implementation made by the LLM. As it will be explained in the following section, the analysis of such a decision will be useful to automate the evaluation process.

For our experiments, we defined one prompt per each of the problems described in Section~\ref{subsec:data_acquisition}. Each prompt incorporated three different sample implementations.

\subsection{LLM Feedback Analysis Automation}\label{subsec:code_analysis_automation}
Provided that the LLM feedback adheres to the fixed structure given by the prompt in Figure~\ref{fig:prompt}, its automated analysis should be straightforward. Since the feedback is provided as a Markdown document, it can be converted into other programmatically-analyzable formats using previously existing libraries. For this work we used the \lstinline{markdown-to-json} Python~package\footnote{\url{https://pypi.org/project/markdown-to-json/}}, which incorporates a function converting a Markdown string into a Python dictionary.

Once the Markdown document is converted into a dictionary, the content of each section can be extracted and analyzed separately. One fundamental item that will be analyzed is the prediction of the LLM regarding the correctness of the implementation. This binary value can be automatically extracted and used to gain knowledge about the LLM ability to understand code. For this purpose, it is important to note that the functions that should be implemented in introductory computer programming assignments can typically be checked by executing a small set of unit tests or asserts. Provided that the set of unit tests covers all relevant cases, the function correctness can be determined automatically and then compared with the LLM correctness prediction. This results in four possible cases that are depicted in Table~\ref{tab:cases}.

\begin{table}[h!]
\caption{Possible cases when analyzing LLM's prediction about the correctness of the implementation.}
\label{tab:cases}
\begin{center}
\begin{scriptsize}
\begin{tabular}{l|l|l|}
\cline{2-3}
                                              & \multicolumn{1}{l|}{\textbf{Predicted Correct}} & \multicolumn{1}{l|}{\textbf{Predicted Incorrect}} \\ \hline
\multicolumn{1}{|l|}{\textbf{Asserts Ok}}     & True Positives (TP)                             & False Negatives (FN)                              \\ \hline
\multicolumn{1}{|l|}{\textbf{Asserts not Ok}} & False Positives (FP)                            & True Negatives (TN)                               \\ \hline
\end{tabular}
\end{scriptsize}
\end{center}
\end{table}

The analysis of the code submitted by the students, including its extraction from Python notebooks, the verification of unit tests, the analysis of the LLM feedback mentioned above, as well as the extraction of statistics, was implemented as a pipeline or workflow, which was executed using specialized workflow execution software called DeBasher~\cite{ortiz25daf}.

\subsection{Research Question Data Analysis}
In this section, the details of the data treatment related to the different research questions are described.

\subsubsection{RQ1} This research question relates to ChatGPT's ability to understand code developed by a student, determining whether it is correct or not. The prompt adopted in this work allows retrieving the prediction made by ChatGPT about the code correctness (see Section~\ref{subsec:prompting_strategy}). The validity of this prediction can be verified by applying a comprehensive set of unit tests or asserts to the student's code. The verification will produce counts informing about how well the model as a code correctness predictor. In particular, the following measures will be reported:
\begin{itemize}
\item {\bf Accuracy}: The proportion of correctly identified cases among the total cases, indicating overall correctness of predictions.
\item {\bf Sensitivity}:  The proportion of true positive cases among all cases where the condition under study is true (in our case, the implementations that pass the asserts).
\item {\bf Specificity}: The proportion of true negatives among all cases where the condition being studied is false (in this paper, the implementations not passing the asserts).
\end{itemize}

\subsubsection{RQ2} For this research question, the corrected version of the student code generated by the LLM is analyzed. Such code can be extracted from the LLM feedback, since it contains a specific section to store it. The extracted code can be used to execute a set of unit tests, determining its correctness. In addition to this, we are also interested to measure how different the corrected version of the code is with respect to the original student version. Higher degrees of similarity means that the LLM generates the corrected version starting from the student's code instead of from scratch. We measure the difference using the Levenshtein distance~\citep{Levenshtein66bcc} at character level between the code versions. This measure is also called character error rate (CER).

\subsubsection{RQ3} The goal of RQ3 is to study whether the LLM feedback points out real issues in the student code. For this purpose, when evaluating an LLM feedback, we will consider two different situations. First, the feedback identifies at least one issue affecting code correctness. Second, the feedback identifies issues uninvolved with code correctness. As a result of this identification, a set of counts will be collected. Counts are only relevant when the system correctly identifies a student code as faulty, that is, for the TN cases shown in Table~\ref{tab:cases}.

\subsubsection{RQ4} This research question studies the ability of the students to detect helpful feedback and distinguish it from that pointing out non-existent or irrelevant code issues. For this part of the evaluation, specific data have been collected, as it is described in Section~\ref{subsec:user_eval}. Such data consisted in the answers to three questions for four problem implementations (two for \probone\ and another two for \probtwo). The first question, \userqone, asked the student whether he/she thinks that the LLM feedback pointed out to real issues in the code being analyzed. The second question, \userqtwo, requested students to provide a score between 1 and 5 evaluating the usefulness of the feedback. The third question, \userqthree, requested the students to explain which information they used to decide about the feedback correctness. For \userqone, counts about correct or incorrect identification of relevant LLM feedback will be collected. For \userqtwo, the student score for the LLM feedbacks will be averaged. Finally, for \userqthree, a qualitative analysis of student's strategies to identify unhelpful feedback will be carried out.

\subsubsection{RQ5} The goal of RQ5 is to determine whether ChatGPT's feedback shows a consistent structure that can be analyzed in a programmatic manner. For this purpose, the model responses will be compared to the intended feedback structure detailed by the used prompt (see Figure~\ref{fig:prompt}). Again, this verification will produce a set of counts that will be reported.

\subsubsection{RQ6} Provided that some of the measurements described for previous research questions can be obtained automatically, in RQ6 we will look for relationships between such automatic measures and other measures that can only be produced manually. The most important manual measure used in this work is the number of times that the LLM feedback for a particular student code points out at least one issue preventing correctness. Such measure will be compared with the accuracy of the LLM as a code correctness predictor, since we see such accuracy as an indicator of the ability of the LLM to understand the student code. Another automatic measure that is potentially related with the previously mentioned manual measure is the ability of the LLM to generate corrected code that passes the asserts, since it is reasonable to assume that no corrected version of a faulty code can be proposed without really understanding the issues relevant to code correctness.

\section{Results}\label{sec:results}
The following sections show the obtained results for each research
question.

\subsection{RQ1}
RQ1 focuses on analyzing ChatGPT's ability to understand code, formalizing such analysis as a binary classification task, where the classifier's prediction is extracted from the LLM feedback (see Section~\ref{subsec:prompting_strategy}) and the ground truth is obtained by executing asserts on the student's code.

Before presenting the classification results, it is of interest to show some statistics about the implementations to be analyzed. Table~\ref{tab:data_general} shows the total implementations for the problems considered in this article. From the total, the absolute and relative frequencies of implementations with runtime exceptions and those that pass or fail the asserts were determined. As can be seen, the number of implementations with runtime exceptions was reduced, and the percentages of implementations that pass or fail the asserts are balanced.

\begin{table}[h!]
\caption{General data about the student implementations of the problems considered in this study, including those with runtime exceptions and those passing or not the asserts.}
\label{tab:data_general}
\begin{center}
\begin{scriptsize}
\begin{tabular}{c|r|r|r|r|}
\cline{2-5}
\multicolumn{1}{l|}{}             & \textbf{Total} & \textbf{Runtime Ex.} & \textbf{Asserts Ok} & \textbf{Asserts not Ok} \\ \hline
\multicolumn{1}{|c|}{\textbf{\probone}}   & 118            & 14 (11.9\%)          & 59 (50.0\%)         & 59 (50.0\%)             \\ \hline
\multicolumn{1}{|c|}{\textbf{\probtwo}}   & 116            & 24 (20.7\%)          & 56 (48.3\%)         & 60 (51.7\%)             \\ \hline
\multicolumn{1}{|c|}{\textbf{\probthree}} & 94             & 6 (6.4\%)            & 54 (57.4\%)         & 40 (42.6\%)             \\ \hline
\multicolumn{1}{|c|}{\textbf{\probfour}}  & 92             & 12 (13.0\%)          & 47 (51.1\%)         & 45 (48.9\%)             \\ \hline
\multicolumn{1}{|c|}{\textbf{\probfive}}  & 94             & 10 (10.6\%)          & 60 (63.8\%)         & 34 (36.2\%)             \\ \hline
\end{tabular}
\end{scriptsize}
\end{center}
\end{table}

Table~\ref{tab:rq1} shows the accuracy, sensitivity and specificity for both \gptthree\ and \gptfour\ and the problems being studied. As can be seen, \gptthree\ achieves moderate accuracy with \probone, \probtwo\ and \probfive, and poor accuracy for \probthree\ and \probfour. We attribute the lower accuracy to the higher difficulty of \probthree\ and \probfour, which were solved by the students using recursion in many cases. On the other hand, \gptfour\ showed greater accuracy than \gptthree\ in all cases, including the two problems, \probthree\ and \probfour, that we identify as more difficult. Both models showed greater specificity than sensitivity, anticipating a greater ability to assist a student whose code does not pass the asserts (although for \gptthree\ and problems \probthree\ and \probfour, this would merely indicate just a strong tendency of the model to classify implementations as incorrect). Given that the LLM feedback will be more useful precisely in those situations where the student has implemented defective code, we believe that specificity is more important than sensitivity.

\begin{table}[h!]
\caption{Accuracy, sensitivity and specificity for both models when applied to the problems considered in this work.}
\label{tab:rq1}
\begin{center}
\begin{scriptsize}
\begin{tabular}{cl|r|r|r|}
\cline{3-5}
\multicolumn{1}{l}{}                               &                   & \textbf{Accuracy} & \textbf{Sensitivity} & \textbf{Specificity} \\ \hline
\multicolumn{1}{|c|}{\multirow{2}{*}{\textbf{\probone}}} & \textbf{GPT-3.5T}   & 67.8\%      & 93.2\%               & 42.4\%       \\ \cline{2-5}
\multicolumn{1}{|c|}{}                             & \textbf{GPT-4T}           & 86.4\%      & 76.2\%               & 96.6\%       \\ \hline
\multicolumn{1}{|c|}{\multirow{2}{*}{\textbf{\probtwo}}} & \textbf{GPT-3.5T}   & 76.7\%      & 82.1\%               & 71.7\%       \\ \cline{2-5}
\multicolumn{1}{|c|}{}                             & \textbf{GPT-4T}           & 85.3\%      & 78.5\%               & 91.7\%       \\ \hline
\multicolumn{1}{|c|}{\multirow{2}{*}{\textbf{\probthree}}} & \textbf{GPT-3.5T} & 54.3\%      & 29.6\%               & 87.5\%       \\ \cline{2-5}
\multicolumn{1}{|c|}{}                             & \textbf{GPT-4T}           & 61.7\%      & 55.6\%               & 70.0\%       \\ \hline
\multicolumn{1}{|c|}{\multirow{2}{*}{\textbf{\probfour}}} & \textbf{GPT-3.5T}  & 52.2\%      & 12.8\%               & 93.3\%       \\ \cline{2-5}
\multicolumn{1}{|c|}{}                             & \textbf{GPT-4T}           & 62.0\%      & 25.5\%               & 100.0\%       \\ \hline
\multicolumn{1}{|c|}{\multirow{2}{*}{\textbf{\probfive}}} & \textbf{GPT-3.5T}  & 60.6\%      & 41.7\%               & 94.1\%       \\ \cline{2-5}
\multicolumn{1}{|c|}{}                             & \textbf{GPT-4T}           & 78.7\%      & 70.0\%               & 94.1\%       \\ \hline
\end{tabular}
\end{scriptsize}
\end{center}
\end{table}

\subsection{RQ2}\label{subsec:results_rq2}
The objective of RQ2 is to study the main characteristics of the corrected version of code generated by the LLM in case the student's implementation was incorrect.

Table~\ref{tab:rq2} shows the percentages related to the corrected versions of student implementations that compile, generate runtime exceptions, and pass the asserts. Additionally, the table also displays a measure of distance between the model's code and the student's code by averaging the CER for each implementation. As can be seen, the corrected versions of both models almost always compiled without any issues. However, the implementations generated by \gptthree\ had runtime errors in a small percentage of cases, and around one-fourth did not pass the specific asserts for each problem. On the other hand, for \gptfour, the implementations had a negligible amount of runtime errors and in almost all cases, they passed the asserts. Finally, no significant differences were observed in the CER for both models, indicating that they construct their solution to a similar extent based on the student's implementation, with the exception of problem \probthree, which showed greater CER scores (a CER value around 50\% means that a half of the characters of the student implementation needed edition in order to generate the corrected version).

\begin{table}[h!]
\caption{Fractions of the corrected versions of the student code that compile, result in runtime exceptions and pass the asserts for the problems considered in this work. A measure of the difference between the student and the LLM code in terms of CER is also shown (the lower, the most similar the LLM code is to the student one).}
\label{tab:rq2}
\begin{center}
\begin{scriptsize}
\begin{tabular}{cl|r|r|r|r|}
\cline{3-6}
\multicolumn{1}{l}{}                               &                   & \textbf{Compile} & \textbf{Runtime Ex.} & \textbf{Asserts Ok} & \textbf{CER} \\ \hline
\multicolumn{1}{|c|}{\multirow{2}{*}{\textbf{\probone}}} & \textbf{GPT-3.5T}   & 100\%            & 13.3\%         & 73.3\%              & 52.2\%       \\ \cline{2-6}
\multicolumn{1}{|c|}{}                             & \textbf{GPT-4T}           & 100\%            & 2.9\%          & 97.0\%              & 54.2\%       \\ \hline
\multicolumn{1}{|c|}{\multirow{2}{*}{\textbf{\probtwo}}} & \textbf{GPT-3.5T}   & 100\%            & 9.0\%          & 74.5\%              & 46.4\%       \\ \cline{2-6}
\multicolumn{1}{|c|}{}                             & \textbf{GPT-4T}           & 97.1\%           & 0.0\%          & 94.1\%              & 54.6\%       \\ \hline
\multicolumn{1}{|c|}{\multirow{2}{*}{\textbf{\probthree}}} & \textbf{GPT-3.5T} & 100\%            & 1.4\%          & 97.3\%              & 77.4\%       \\ \cline{2-6}
\multicolumn{1}{|c|}{}                             & \textbf{GPT-4T}           & 100\%            & 0.0\%          & 100.0\%             & 70.3\%       \\ \hline
\multicolumn{1}{|c|}{\multirow{2}{*}{\textbf{\probfour}}} & \textbf{GPT-3.5T}  & 100\%            & 10.7\%         & 92.9\%              & 44.3\%       \\ \cline{2-6}
\multicolumn{1}{|c|}{}                             & \textbf{GPT-4T}           & 100\%            & 2.4\%          & 98.8\%              & 55.2\%       \\ \hline
\multicolumn{1}{|c|}{\multirow{2}{*}{\textbf{\probfive}}} & \textbf{GPT-3.5T}  & 100\%            & 1.5\%          & 97.1\%              & 50.3\%       \\ \cline{2-6}
\multicolumn{1}{|c|}{}                             & \textbf{GPT-4T}           & 100\%            & 1.5\%          & 98.5\%              & 62.5\%       \\ \hline
\end{tabular}
\end{scriptsize}
\end{center}
\end{table}

\subsection{RQ3}

RQ3 focuses on analyzing the feedback from the LLM, including those indicating issues that cause the code to be incorrect and others pointing out issues that do not affect the correctness of the code, either because such problems are not actually relevant to passing the unit tests (e.g. using a \lstinline{while} loop when a \lstinline{for} loop could have been used) or because they correspond to non-existent problems (e.g. stating that the code generates incorrect output for a given input when it does not).

Table~\ref{tab:rq3} displays the issues found in the LLM feedback for all implementations of the five problems under study that did not pass the unit tests. Specifically, it shows the fraction of feedbacks containing one or more actual problems in the code, as well as the number of feedbacks mentioning issues uninvolved in passing unit tests. As can be observed, \gptthree\ has poor ability to identify one or more problems in the code of problem \probone, moderate ability when it comes to problem \probtwo, and good ability for the rest. In contrast, \gptfour\ was able to identify at least one problem in over 70\% of the feedbacks for the five problems. However, \gptfour\ showed a higher tendency to point out issues in the code unrelated to passing the unit tests, affecting over 45\% of the feedbacks for the five problems.

\begin{table}[h!]
  \caption{For all implementations of problems not passing the asserts, the table shows the fraction of feedbacks for models \gptthree\ and \gptfour\ indicating one or more issues involved in passing unit tests, or indicating unrelated problems.}
\label{tab:rq3}
\begin{center}
\begin{scriptsize}
\begin{tabular}{ll|rr|}
\cline{3-4}
                                                   &                   & \multicolumn{2}{c|}{\textbf{Issues Detected by the LLM}}        \\ \cline{3-4}
                                                   &                   & \multicolumn{1}{c|}{\textbf{One or More}} & \multicolumn{1}{c|}{\textbf{Uninvolved}} \\ \hline
\multicolumn{1}{|l|}{\multirow{2}{*}{\textbf{\probone}}} & \textbf{GPT-3.5T} & \multicolumn{1}{r|}{37.3\%}         & 11.9\%              \\ \cline{2-4}
\multicolumn{1}{|l|}{}                             & \textbf{GPT-4T}   & \multicolumn{1}{r|}{88.1\%}               & 52.5\%              \\ \hline
\multicolumn{1}{|l|}{\multirow{2}{*}{\textbf{\probtwo}}} & \textbf{GPT-3.5T} & \multicolumn{1}{r|}{56.7\%}         & 30.0\%              \\ \cline{2-4}
\multicolumn{1}{|l|}{}                             & \textbf{GPT-4T}   & \multicolumn{1}{r|}{88.3\%}               & 51.6\%              \\ \hline
\multicolumn{1}{|l|}{\multirow{2}{*}{\textbf{\probthree}}} & \textbf{GPT-3.5T} & \multicolumn{1}{r|}{75.0\%}       & 40.0\%              \\ \cline{2-4}
\multicolumn{1}{|l|}{}                             & \textbf{GPT-4T}   & \multicolumn{1}{r|}{67.5\%}               & 47.5\%              \\ \hline
\multicolumn{1}{|l|}{\multirow{2}{*}{\textbf{\probfour}}} & \textbf{GPT-3.5T} & \multicolumn{1}{r|}{75.0\%}        & 35.6\%              \\ \cline{2-4}
\multicolumn{1}{|l|}{}                             & \textbf{GPT-4T}   & \multicolumn{1}{r|}{95.6\%}               & 51.1\%                  \\ \hline
\multicolumn{1}{|l|}{\multirow{2}{*}{\textbf{\probfive}}} & \textbf{GPT-3.5T} & \multicolumn{1}{r|}{76.5\%}        & 26.5\%              \\ \cline{2-4}
\multicolumn{1}{|l|}{}                             & \textbf{GPT-4T}   & \multicolumn{1}{r|}{91.2\%}               & 45.4\%              \\ \hline
\end{tabular}
\end{scriptsize}
\end{center}
\end{table}

Finally, from those feedbacks pointing out issues unrelated to code correctness for incorrect programs, we were interested in obtaining the fraction that mentioned non-existing issues. Results are shown in Table~\ref{tab:rq3_nonexist}. As can be seen in the table, around a half of the \gptthree\ feedbacks incorporated non-existent issues. For the \gptfour\ model, the fractions were lower but not insignificant.

\begin{table}[h!]
  \caption{Given the set of LLM feedbacks indicating issues not related to code correctness for incorrect programs, the table shows the fraction of feedbacks from that set mentioning non-existent issues.}
\label{tab:rq3_nonexist}
\begin{center}
\begin{scriptsize}
\begin{tabular}{ll|r|}
\cline{3-3}
                                                   &                   & \textbf{Non-Existent Issues} \\ \hline
\multicolumn{1}{|l|}{\multirow{2}{*}{\textbf{\probone}}} & \textbf{GPT-3.5T}    & 85.7\%                       \\ \cline{2-3}
\multicolumn{1}{|l|}{}                             & \textbf{GPT-4T}            & 61.3\%                       \\ \hline
\multicolumn{1}{|l|}{\multirow{2}{*}{\textbf{\probtwo}}} & \textbf{GPT-3.5T}    & 88.3\%                       \\ \cline{2-3}
\multicolumn{1}{|l|}{}                             & \textbf{GPT-4T}            & 38.7\%                       \\ \hline
\multicolumn{1}{|l|}{\multirow{2}{*}{\textbf{\probthree}}} & \textbf{GPT-3.5T}  & 93.8\%                       \\ \cline{2-3}
\multicolumn{1}{|l|}{}                             & \textbf{GPT-4T}            & 78.9\%                       \\ \hline
\multicolumn{1}{|l|}{\multirow{2}{*}{\textbf{\probfour}}} & \textbf{GPT-3.5T}   & 87.5\%                       \\ \cline{2-3}
\multicolumn{1}{|l|}{}                             & \textbf{GPT-4T}            & 65.2\%                       \\ \hline
\multicolumn{1}{|l|}{\multirow{2}{*}{\textbf{\probfive}}} & \textbf{GPT-3.5T}   & 66.7\%                       \\ \cline{2-3}
\multicolumn{1}{|l|}{}                             & \textbf{GPT-4T}            & 93.3\%                       \\ \hline
\end{tabular}
\end{scriptsize}
\end{center}
\end{table}

\subsection{RQ4}
Our LLM feedback evaluation would not be complete without the involvement of the potential users of a hypothetical LLM-based learning tool. For this reason, RQ4 examines the feedback from this perspective. Due to the fact that \gptfour\ showed better performance than \gptthree, it was decided to focus this experimentation solely on the former model.

Table~\ref{tab:rq4} assesses the students' ability to correctly predict if a particular feedback provided by \gptfour\ indicated actually existing errors in a set consisting of four implementations: cases \emph{a} and \emph{b} for \probone\ and the same for \probtwo\ (see Section~\ref{subsec:user_eval} for a more detailed description). Additionally, the table also displays whether the feedback was actually correct, the number of responses, the number of times the student considered the feedback correct (this corresponds to question Q1 of the survey), and finally, the average score from 1 to 5 regarding the degree of usefulness students found for each feedback, with 1 corresponding to ``very little useful'' and 5 to ``very useful''. Only those students answering affirmatively to Q1 could provide a usefulness score.

As can be seen in Table~\ref{tab:rq4}, the accuracy when determining feedback correctness was only high for case \emph{a} of \probone. In that case, the implementation to be analyzed was very simple and there was only one issue affecting code correctness. The situation was different for the rest of the implementations that were analyzed by the students. Interestingly, for case \emph{b} of \probone\ and \probtwo, where ChatGPT did not provide correct feedback, there were students who were convinced that the issues pointed out by the LLM were accurate. In those cases, the usefulness score provided by the students was equal or above 3 points. The usefulness score was higher for case \emph{a} of both problems, where ChatGPT was providing correct feedback. In general, the results shown in Table~\ref{tab:rq4} suggest that the students involved in the study struggled to determine whether a given LLM feedback was correct, unless such feedback and the code to be analyzed were simple.

\begin{table}[h!]
\caption{Student's ability in terms of accuracy to correctly determine whether a given \gptfour\ feedback pointed out existent issues in a set of four implementations of problems (cases \emph{a} and \emph{b} for \probone\ and the same for \probtwo). The table also shows whether the feedback was correct or not for each case, the number of student responses, the number of students that considered such feedback correct (this was asked in question Q1 of the questionnaire), and for the affirmative responses, the average of the usefulness score given by the students, a value of 1 means ``very little useful'' and a value of 5 means ``very useful''.}
\label{tab:rq4}
\begin{center}
\begin{scriptsize}
\begin{tabular}{ll|r|r|r|r|r|}
\cline{3-7}
                                                   &                        & \textbf{LLM Feedback Correct} & \textbf{Responses} & \textbf{Q1 Affirmative} & \textbf{Accuracy} & \textbf{Usefulness} \\ \hline
\multicolumn{1}{|l|}{\multirow{2}{*}{\textbf{P1}}} & \textbf{Case \emph{a}} & Yes       & 11 & 11           & 100\%             & 4.4                 \\ \cline{2-7}
\multicolumn{1}{|l|}{}                             & \textbf{Case \emph{b}} & No        & 9  & 3            & 66.7\%            & 4.0                 \\ \hline
\multicolumn{1}{|l|}{\multirow{2}{*}{\textbf{P2}}} & \textbf{Case \emph{a}} & Yes       & 10 & 5            & 50.0\%            & 3.0                 \\ \cline{2-7}
\multicolumn{1}{|l|}{}                             & \textbf{Case \emph{b}} & No        & 10 & 6            & 40.0\%            & 3.7                 \\ \hline
\end{tabular}
\end{scriptsize}
\end{center}
\end{table}

Additionally, the students were also asked about what factors influenced them when determining if the feedback indicated real errors in the code. As mentioned before, they were not always able to identify inaccurate feedback. In many cases, the students that successfully identified whether the feedback was correct or not, executed the code following the feedback statements. There were also situations where an erroneous feedback just seemed ``perfect'' to the students, suggesting that ChatGPT was being considered as some sort of authority figure. On the contrary, there were students who did not trust the model feedback from the very beginning because it seemed confusing or difficult to understand and preferred to diagnose the errors in the implementation by themselves. Interestingly, some students mistrusted ChatGPT's feedback because it did not include code. Descriptive responses were also positively valued by the students. Finally, in other occasions, some feedback statements were judged as correct and others as incorrect in spite of the fact that the four feedbacks provided to the students did not mix correct and incorrect statements. This situation clearly reflected the difficulties of the students to work with LLM feedback.

\subsection{RQ5}
The ability to automatically evaluate ChatGPT's suitability as a teaching assistant depends on the response to RQ5, whose goal was to verify whether the structure of the LLM feedback was consistent.

Table~\ref{tab:rq5} shows, for both models and problems \probone\ to \probfive, the number of feedbacks containing code outside the predefined section for it, the number of feedbacks missing any of the sections established by the prompt (see Section~\ref{subsec:prompting_strategy}), those incorporating extra sections, and finally, those containing a correct structure without missing or extra sections. As can be observed, \gptthree's performance in this aspect was outstanding, with percentages of feedback with strictly correct structures exceeding 96\%. The situation was not as good for \gptfour, mainly due to the fact that it often added extra sections. However, missing sections were very infrequent.

\begin{table}[h!]
\caption{Fractions of LLM feedbacks for \gptthree\ and \gptfour\ models and the problems considered in this work that contained misplaced code, missing sections, extra sections or a correct structure (no missing or extra sections).}
\label{tab:rq5}
\begin{center}
\begin{scriptsize}
\begin{tabular}{cl|r|r|r|r|}
\cline{3-6}
\multicolumn{1}{l}{}                               &                   & \textbf{Misplaced Code} & \textbf{Missing Sections} & \textbf{Extra Sections} & \textbf{Correct Structure} \\ \hline
\multicolumn{1}{|c|}{\multirow{2}{*}{\textbf{\probone}}} & \textbf{GPT-3.5T}   & 0.0\%       & 0.0\%     & 0.8\%       & 99.2\%          \\ \cline{2-6}
\multicolumn{1}{|c|}{}                             & \textbf{GPT-4T}           & 5.9\%       & 0.8\%     & 23.7\%      & 75.4\%          \\ \hline
\multicolumn{1}{|c|}{\multirow{2}{*}{\textbf{\probtwo}}} & \textbf{GPT-3.5T}   & 0.9\%       & 3.4\%     & 0.0\%       & 96.6\%          \\ \cline{2-6}
\multicolumn{1}{|c|}{}                             & \textbf{GPT-4T}           & 3.4\%       & 6.0\%     & 36.2\%      & 57.8\%          \\ \hline
\multicolumn{1}{|c|}{\multirow{2}{*}{\textbf{\probthree}}} & \textbf{GPT-3.5T} & 0.0\%       & 1.0\%     & 0.0\%       & 98.9\%          \\ \cline{2-6}
\multicolumn{1}{|c|}{}                             & \textbf{GPT-4T}           & 6.4\%       & 0.0\%     & 27.7\%      & 72.3\%          \\ \hline
\multicolumn{1}{|c|}{\multirow{2}{*}{\textbf{\probfour}}} & \textbf{GPT-3.5T}  & 0.0\%       & 1.1\%     & 0.0\%       & 98.9\%          \\ \cline{2-6}
\multicolumn{1}{|c|}{}                             & \textbf{GPT-4T}           & 2.2\%       & 3.3\%     & 19.6\%      & 76.1\%          \\ \hline
\multicolumn{1}{|c|}{\multirow{2}{*}{\textbf{\probfive}}} & \textbf{GPT-3.5T}  & 0.0\%       & 1.1\%     & 0.0\%       & 98.9\%          \\ \cline{2-6}
\multicolumn{1}{|c|}{}                             & \textbf{GPT-4T}           & 9.6\%       & 2.1\%     & 57.4\%      & 40.4\%          \\ \hline
\end{tabular}
\end{scriptsize}
\end{center}
\end{table}

Thanks to the fact that the feedback structures were always programmatically analyzable and almost always incorporated all of the sections specified in the problem prompt, it was possible to automate a large portion of the results obtained up to this point in the article. In particular, all the results from tables~\ref{tab:rq1},~\ref{tab:rq2} and \ref{tab:rq5} were automatically obtained. Moreover, in the next section, additional automatic results useful to evaluate LLM feedback will be shown.

\subsection{RQ6}
The purpose of the last research question is to determine if there are relationships between manual and automatic evaluation measures, in case the latter exist. The existence of strong relationships between automatic and manual measures could reduce the need for costly manual evaluation processes.

As indicated in the previous section, it has been possible to automate several evaluation measures, particularly those related to research questions RQ1 and RQ2. On the other hand, the evaluation measures obtained within research questions RQ3 and RQ4 correspond to manual measures. We will not study the relationship between the measures of RQ1 and RQ2 with those of RQ4, as we consider that the interest in acquiring the latter is not influenced by the existence of automatic measures. This is because the measures of RQ4 collect the criteria of end-users of a hypothetical learning tool based on LLMs, and those criteria are very difficult to simulate or predict.

However, we will study the relationship between the automatic evaluation measures of RQ1 and RQ2 and those obtained for RQ3. The measures of RQ3 evaluate the model's ability to identify problems in defective code. In this regard, it is interesting to consider the cases presented in Table~\ref{tab:cases}, where four situations are shown based on whether the unit tests are passed or not and whether the model predicts that the implementation under analysis is correct or incorrect. Under these circumstances, it is important to highlight that the model will only produce feedback when it considers the implementation to be incorrect. If we focus on implementations not passing the asserts, we will know from the outset that the feedback is erroneous when the LLM classifies the code as correct (FPs), since it will not point out any code issues. It is only when the program is incorrect and the LLM identifies it as such (TNs) that the feedback will need to be evaluated manually.

If we consider the situation where ChatGPT generates feedback that does not indicate any problem in faulty code, in other words, the complement of the ``One or More'' column of Table~\ref{tab:rq3}, then we have a measure of the rate of erroneous feedbacks made by the model. Due to the fact that the feedback will always be wrong for the FP cases, it is possible to automatically provide a lower bound for such rate. Table~\ref{tab:rq6_1} shows, for the two models and the five problems under study, the percentage of feedbacks requiring manual evaluation with respect to the total. Additionally, for the defective implementations, the percentage of erroneous feedbacks, and the lower bound of this percentage calculated automatically. For both models, less than half of the generated feedbacks required manual evaluation. Regarding the error measures, for \gptthree, both the error rate and its lower bound were high for problems \probone, \probtwo\ and \probthree, to the extent that only considering the lower bound could discard the use of the model without the need for manual evaluation, saving time and effort. In contrast, the lower bound was reduced for \probfour\ and \probfive, making manual evaluation necessary. Regarding, \gptfour, it showed lower values for both measures with respect to \gptthree. In addition to this, the lower bound for the percentage of erroneous feedbacks was always quite low, introducing again the necessity of manually evaluating the output.

\begin{table}[h!]
\caption{Fraction of feedbacks requiring manual evaluation. For all implementations of problems not passing the asserts, fraction of erroneous feedbacks and the lower bound for such fraction when using the \gptthree\ and \gptfour\ models.}
\label{tab:rq6_1}
\begin{center}
\begin{scriptsize}
\begin{tabular}{llr|rr|}
\cline{4-5}
                                                   &  & & \multicolumn{2}{c|}{\textbf{Fraction of Model Feedbacks}} \\ \cline{3-5}
                                                   & \multicolumn{1}{l|}{} & \multicolumn{1}{l|}{\textbf{Manual Eval.}} & \multicolumn{1}{l|}{\textbf{Erroneous}} & \textbf{Erroneous (Lower Bound)} \\ \hline
\multicolumn{1}{|l|}{\multirow{2}{*}{\textbf{\probone}}} & \multicolumn{1}{l|}{\textbf{GPT-3.5T}}  & \multicolumn{1}{r|}{21.2\%} & \multicolumn{1}{r|}{62.7\%} & 57.6\%  \\ \cline{2-5}
\multicolumn{1}{|l|}{}                             & \multicolumn{1}{l|}{\textbf{GPT-4T}}          & 48.3\%                      & \multicolumn{1}{r|}{11.8\%} & 3.4\%   \\ \hline
\multicolumn{1}{|l|}{\multirow{2}{*}{\textbf{\probtwo}}} & \multicolumn{1}{l|}{\textbf{GPT-3.5T}}  & \multicolumn{1}{r|}{37.1\%} & \multicolumn{1}{r|}{43.3\%} & 28.3\%  \\ \cline{2-5}
\multicolumn{1}{|l|}{}                             & \multicolumn{1}{l|}{\textbf{GPT-4T}}          & 47.4\%                      & \multicolumn{1}{r|}{11.7\%} & 8.3\%   \\ \hline
\multicolumn{1}{|l|}{\multirow{2}{*}{\textbf{\probthree}}} & \multicolumn{1}{l|}{\textbf{GPT-3.5T}}& \multicolumn{1}{r|}{37.2\%} & \multicolumn{1}{r|}{25.0\%} & 10.0\%  \\ \cline{2-5}
\multicolumn{1}{|l|}{}                             & \multicolumn{1}{l|}{\textbf{GPT-4T}}          & 29.8\%                      & \multicolumn{1}{r|}{32.5\%} & 30.0\%   \\ \hline
\multicolumn{1}{|l|}{\multirow{2}{*}{\textbf{\probfour}}} & \multicolumn{1}{l|}{\textbf{GPT-3.5T}} & \multicolumn{1}{r|}{45.7\%} & \multicolumn{1}{r|}{24.4\%} & 4.4\%  \\ \cline{2-5}
\multicolumn{1}{|l|}{}                             & \multicolumn{1}{l|}{\textbf{GPT-4T}}          & 48.9\%                      & \multicolumn{1}{r|}{4.4\%}  & 0.0\%   \\ \hline
\multicolumn{1}{|l|}{\multirow{2}{*}{\textbf{\probfive}}} & \multicolumn{1}{l|}{\textbf{GPT-3.5T}} & \multicolumn{1}{r|}{34.0\%} & \multicolumn{1}{r|}{20.6\%} & 2.9\%  \\ \cline{2-5}
\multicolumn{1}{|l|}{}                             & \multicolumn{1}{l|}{\textbf{GPT-4T}}          & 34.0\%                      & \multicolumn{1}{r|}{5.9\%}  & 5.9\%   \\ \hline
\end{tabular}
\end{scriptsize}
\end{center}
\end{table}

It is important to highlight the relationship between the automatic specificity measure for RQ1 presented in Table~\ref{tab:rq1} and the measures presented in Table~\ref{tab:rq6_1} just shown. In particular, the lower bound of the rate of erroneous feedbacks is the complement of the specificity, also known as the false positive rate (FPR). A low specificity results in a higher number of FP cases, and therefore the rate of erroneous feedbacks and its lower bound will be higher. This situation would be similar to that observed for the \gptthree\ model and problems \probone\ to \probthree. On the contrary, a model with a high specificity generates a large number of TNs that need to be evaluated manually, reducing the lower bound of the rate of erroneous feedbacks. This would correspond to the situation observed for the \gptfour\ model.

On the other hand, it is of interest to study whether there is any relationship between feedback quality and the ability of the LLM to generate a corrected version of the student's faulty code, an ability that was analyzed for RQ2 (see Section~\ref{subsec:results_rq2}). It could be hypothesized that if the system can generate a corrected version passing the asserts from the student's code, then it should have successfully identified the existing errors, resulting in better feedback. To test this hypothesis, Table~\ref{tab:rq6_2} shows the feedback mentioning at least one error related to code correction and the feedback mentioning uninvolved problems based on whether the corrected version (CV) of the code passed the unit tests or not. The experiment was conducted only for the \gptthree\ model, as it was the only one where there was a certain percentage of corrected versions that did not pass the asserts. As can be seen, the initial hypothesis is not confirmed, since the fraction of feedbacks mentioning at least one problem related to code correction when the asserts were not passed increased instead of decreased in most cases. Regarding the fraction of feedbacks containing problems unrelated to code correction, it increased or decreased when the asserts were not passed, depending on the problem being considered. In general, the results suggest a possible disconnection between the process of generating the corrected version of the student code and the rest of the analysis carried out by the LLM.

\begin{table}[h!]
\caption{Fraction of feedbacks mentioning at least one issue affecting code correctness and fraction of feedbacks mentioning uninvolved issues for the problems considered in this work depending on whether the corrected version (CV) generated by \gptthree\ passed the asserts or not.}
\label{tab:rq6_2}
\begin{center}
\begin{scriptsize}
\begin{tabular}{ll|rr|}
\cline{3-4}
                                                   &                         & \multicolumn{2}{l|}{\textbf{Issues Detected by the LLM}}        \\ \cline{3-4}
                                                   &                         & \multicolumn{1}{l|}{\textbf{One or More}} & \textbf{Uninvolved} \\ \hline
\multicolumn{1}{|l|}{\multirow{2}{*}{\textbf{\probone}}} & \textbf{Asserts Ok for CV}     & \multicolumn{1}{r|}{72.7\%}  & 31.8\%                     \\ \cline{2-4}
\multicolumn{1}{|l|}{}                             & \textbf{Asserts Not Ok for CV}       & \multicolumn{1}{r|}{75.0\%}  & 50.0\%                    \\ \hline
\multicolumn{1}{|l|}{\multirow{2}{*}{\textbf{\probtwo}}} & \textbf{Asserts Ok for CV}     & \multicolumn{1}{r|}{60.9\%}  & 51.2\%                    \\ \cline{2-4}
\multicolumn{1}{|l|}{}                             & \textbf{Asserts Not Ok for CV}       & \multicolumn{1}{r|}{71.4\%}  & 50.0\%                    \\ \hline
\multicolumn{1}{|l|}{\multirow{2}{*}{\textbf{\probthree}}} & \textbf{Asserts Ok for CV}   & \multicolumn{1}{r|}{38.9\%}  & 73.6\%                    \\ \cline{2-4}
\multicolumn{1}{|l|}{}                             & \textbf{Asserts Not Ok for CV}       & \multicolumn{1}{r|}{100.0\%} & 50.0\%                    \\ \hline
\multicolumn{1}{|l|}{\multirow{2}{*}{\textbf{\probfour}}} & \textbf{Asserts Ok for CV}    & \multicolumn{1}{r|}{41.0\%}  & 66.7\%                    \\ \cline{2-4}
\multicolumn{1}{|l|}{}                             & \textbf{Asserts Not Ok for CV}       & \multicolumn{1}{r|}{33.3\%}  & 83.3\%                    \\ \hline
\multicolumn{1}{|l|}{\multirow{2}{*}{\textbf{\probfive}}} & \textbf{Asserts Ok for CV}    & \multicolumn{1}{r|}{36.3\%}  & 66.7\%                    \\ \cline{2-4}
\multicolumn{1}{|l|}{}                             & \textbf{Asserts Not Ok for CV}       & \multicolumn{1}{r|}{100.0\%} & 0.0\%                    \\ \hline
\end{tabular}
\end{scriptsize}
\end{center}
\end{table}

\section{Discussion}\label{sec:discussion}
The following sections are devoted to discuss the three article goals specified in the introductory section: evaluating the performance of ChatGPT as a teaching assistant (Section~\ref{subsec:chatgpt_performance}), programmatically analyzing ChatGPT's feedback (Section~\ref{subsec:chatgpt_autom}) and proposing a way to operate ChatGPT within a practical application (Section~\ref{subsec:eff_oper}).

\subsection{Performance of ChatGPT as a Teaching Assistant}\label{subsec:chatgpt_performance}
Regarding the use of ChatGPT as a teaching assistant, its capability was evaluated in various dimensions by providing feedback for five problems, from \probone\ to \probfive, using the models \gptthree\ and \gptfour. Substantial differences were found between them. Specifically, \gptfour\ proved to be clearly superior to \gptthree\ in determining whether the student's code was correct or not (with accuracies above 60\% and as high as 86.4\% using \gptfour\ compared to those around 50 or 60\% with \gptthree), in generating corrected versions of the code that could pass the predefined unit tests (\gptfour\ generated corrected versions that passed the asserts in more that a 90\% of the cases for all problems, while \gptthree\ obtained less uniform results, including rates of 73.3\% and 74.5\% for problems \probone\ and \probtwo), and in identifying issues actually present in the code under analysis (\gptfour\ identified at least one error affecting the correctness of the code in at least a 67.5\% and as high as a 91.2\% of the occasions depending on the problem being considered, with \gptthree\ achieving at most a 76.5\% performance).

In light of the results, the \gptthree\ model appears to be far from possessing sufficient ability to function as a teaching assistant in the context being considered. Regarding the \gptfour\ model, despite offering a more solid foundation, it also could not be used without exposing the students to the risk of receiving misleading or irrelevant information regarding key aspects affecting code correctness. This problem is particularly important in the context studied in the article, as the students' ability to detect erroneous information in LLM feedback is still developing. The article incorporates a study with potential end users of the system to investigate this issue. The results of this study demonstrate that these users are not always able to correctly identify situations in which certain feedback points out issues unrelated to code correctness.

\subsection{Automating ChatGPT's Evaluation}\label{subsec:chatgpt_autom}
Regarding the possibility of automating evaluation, the use of a prompt incorporating ICL techniques was key in obtaining responses from ChatGPT with a consistent structure that could be programmatically analyzed. This capability was combined with the inclusion of diagnostic information about the model's performance, specifically, its decision on whether the code was correct or not. Since the accuracy of that decision can be verified by executing unit tests on the code under analysis, it was possible to automatically obtain all results related to ChatGPT's ability to understand code (RQ1) and its ability to generate corrected versions of faulty code (RQ2).

The semi-automatic evaluation presented in this article provides interesting advantages compared to conventional manual evaluation. The adopted approach allows, given an implementation problem, to automatically establish a lower bound for the rate of erroneous feedback generated by an LLM. This lower bound would have been useful, for example, to immediately discard the use of \gptthree\ in three of the five problems considered in this study without the need for subsequent manual evaluation.

The information provided by the automatic measures proposed in this work is interesting because it can be used to easily analyze multiple factors involved in the implementation of LLM-based computer programming teaching assistants, including the particular LLM to be used, the implementation problem proposed to the students, the prompt (provided that such prompt uses ICL techniques to enforce an appropriate feedback structure that incorporates the model decision about code correctness for the implementation being analyzed) or even multiple runs of the same prompt (given the stochastic nature of the LLM output).

\subsection{Effective Operation of ChatGPT as a Teaching Assistant}\label{subsec:eff_oper}
The automatic handling of LLM feedback has applications not only in model evaluation but could also be useful when operating the model as an ingredient of a real software application. Whenever a student needs help with a particular implementation, only the part of the feedback explaining the code errors could be extracted and shown to the student. In this way, the possibility of showing the student LLM feedback containing the code implementing the solution would be completely avoided. Other works focused on a context similar to ours have already pointed out the tendency of ChatGPT to incorporate code, without finding a solution. For example, in~\citep{hellas23etr}, it is mentioned that the generated feedback almost always contained code, and in \citep{macneil23dle}, a similar situation is detailed in which the feedback includes code even though the used prompt explicitly requested otherwise.

Accessing the structured content of the LLM would offer additional opportunities. For instance, since the error description is structured as a list of items, it would be possible to show only some of them, so as to favor a gradual construction of the solution by the student (as was mentioned in Section~\ref{subsec_char_good_llm_feedback}). Another content with pedagogic interest that could be shown to the student would be the brief description of the code generated by the LLM, which could help the student to understand what his/her code is exactly doing.

Additionally, automatic feedback handling could also be useful for the early diagnosis of errors in the feedback itself. To do this, it would be necessary to extract from the feedback the prediction of whether the code being analyzed is correct (see Section~\ref{subsec:prompting_strategy}), as well as to run a set of appropriate unit tests on the code being analyzed. From this information, if the student requested help for correct code (because no asserts have been provided, or these do not cover all the cases of interest), it would be possible to detect the FN cases (see Table~\ref{tab:cases}). Similarly, if the student requested help for incorrect code, the FP cases could be detected as well. That is, for both the FN and FP cases, the situation where the system shows incorrect feedback to the student would be prevented.

The only feedback that could not be treated trivially in an automated manner would be the TN cases. In these situations, the system feedback would contain a description of the errors identified in the incorrect code. However, this description would not necessarily have to be accurate, requiring manual evaluation to verify such correctness.

Interestingly, even in these situations, the automated treatment of LLM feedback could also be useful, providing information that could perhaps be used to build some sort of predictor that offers an estimation of feedback quality. For this purpose, one possible source of inspiration to address the problem could be a discipline in the area of machine translation (MT) called quality estimation~\citep{specia09ets}. MT is the process of translating text from one language to another using computer software (the original language is called source language and the destination language is called the target language). The discipline of quality estimation deals with predicting the quality of translations generated by a translation system. For this purpose, a predictor whose parameters have been trained from features extracted from the source and target sentences is built. The trained predictor is then used to predict translation quality measures for the system translations.

In the case of using an LLM as a teaching assistant, features useful for quality estimation could be extracted from the different elements that make up the feedback. Examples of such features could be the sentence embeddings~\citep{reimers19sbs} extracted from the textual sections of the LLM feedback, or the code embeddings~\citep{alon18cld} generated from the code being analyzed. These features would be used to predict whether a given feedback or one of its elements mentions real implementation issues affecting correctness. Prior to this, the predictor would have been trained with manually-labeled samples.

We plan to explore these ideas for feedback quality estimation as future work.

\section{Related Work}\label{sec:related_work}
The articles that explore the applications of LLMs in an educational context can be classified into two groups: those that work with LLMs specialized in coding and those that work with general-purpose LLMs. LLMs specialized in coding include, for example,
CodeBERT~\citep{feng20cap}, CodeT5~\citep{wang21cia}, WizardCoder~\citep{luo23wec} or StarCoder\citep{li23smt}. On the other hand, general-purpose LLMs include, in addition to ChatGPT, a wide range of highly heterogeneous models, such as the Llama~\citep{touvron23loa} and Llama 2~\citep{touvron23lof} family of models developed by Meta, or the LaMDA model~\citep{thoppilan22llm} developed by Alphabet.

Works studying LLMs specialized in coding include~\citep{chen21ell}, where Codex, a GPT language model fine-tuned on publicly available code from GitHub is compared to other code-focused LLMs when used in code generation tasks. Additionally, in~\citep{yuan23eit}, instruction-tuned and fine-tuned LLMs are evaluated in comprehension and code generation tasks. In this case, the study considers both general purpose (e.g. Llama) and code specific (e.g. StarCoder) models.

On the other hand, articles centered on general purpose LLMs encompass evaluations in different settings. For instance, in~\citep{guo23hci}, the performance of ChatGPT is evaluated in fields such as legal, medical or financial. Text summaries created with ChatGPT within the medical domain are analyzed in~\citep{gao23csa}. In~\citep{kabir23wai}, the responses provided by ChatGPT are analyzed in comparison to those from Stack Overflow. Additionally, some studies focus on studying the current limitations of ChatGPT~\citep{borji2023aca,kocon23cjo,mitrovic23coh}, finding that the model frequently generate inaccurate information. ChatGPT alternatives such as Llama or LaMDA were evaluated in their foundational papers~\citep{touvron23loa,touvron23lof,thoppilan22llm}.

\begin{sloppypar}
Within an educational context, several works can be found that deal with evaluating LLMs. For instance, in~\citep{jacobsen23tpa}, the quality of the feedback provided by ChatGPT in the design of learning objectives is evaluated, testing different prompts for this purpose. In~\citep{laskar23ass}, ChatGPT is evaluated alongside other LLMs in a series of academic benchmarks. Similarly, in~\cite{savelka23cgp}, it was studied whether ChatGPT was capable of passing university programming course exams. On the other hand, there are also studies focused on working with code. In~\citep{jalil23cas}, the effectiveness of ChatGPT in teaching software testing courses is evaluated. In~\citep{macneil23dle}, GPT-3 and GPT-4 are applied to detect logical errors in three code examples, and their performance is compared with that of beginner students. In~\citep{hellas23etr}, the ability of the Codex and ChatGPT models to provide feedback to pre-university students on simple programming problems is studied. More recently, in~\citep{azaiz24fgf}, authors conduct work similar to that presented in~\citep{hellas23etr}, but using GPT-4 to provide feedback for programming tasks in a university context.
\end{sloppypar}

From the studies mentioned earlier, those presented in~\citep{hellas23etr} and~\citep{azaiz24fgf} would be the most related to ours. Nevertheless, there are significant differences.  The most important one is that our work is not only focused on carrying out an evaluation of ChatGPT to generate feedback on code, but also on how to programmatically analyze said feedback, making it possible to define automatic evaluation measures, which results in cheaper, faster and more objective evaluation. Moreover, the automated feedback analysis technique that has been proposed also constitutes the foundation of a possible way to operate ChatGPT or other LLMs as computer programming teaching assistants in real scenarios.

\begin{sloppypar}
In addition to this, our work also differs from that presented in~\citep{hellas23etr} and~\citep{azaiz24fgf} in other aspects, such as the programming language or the LLMs used. Specifically, in~\citep{hellas23etr}, the Dart programming language is used, which is a less widespread language than Python. In contrast, the Java programming language is used in~\citep{azaiz24fgf}. Both Java and Python are languages typically used in university programming courses, while Dart may be more appropriate for the pre-university context explored in~\citep{hellas23etr}. Regarding the LLMs studied, in~\citep{hellas23etr}, two LLMs, Codex and GPT 3, are analyzed, of which Codex is discontinued. On the other hand, only one model, GPT-4, is adopted in~\citep{azaiz24fgf}.
\end{sloppypar}

\section{Conclusions}\label{sec:conclusions}
This article has focused on the use of ChatGPT, specifically the \gptthree\ and \gptfour\ models, as teaching assistants in the context of university introductory programming courses. Within this context, three aspects were studied: the quality of the generated feedback, the possibilities for automating evaluation, and finally, how an LLM-based learning tool could be implemented in practice.

Based on the empirical results, we conclude that \gptthree\ is not ready to be used within the context studied in the article. On the other hand, although \gptfour\ showed superior performance, it cannot be ruled out that the feedback it generates may contain misleading information or information that is not directly related to the problems present in the code under analysis. This problem is particularly important provided that a small study we conducted with potential end users of a hypothetical LLM-based learning tool demonstrates that they cannot always detect these situations.

Furthermore, the application of ICL techniques to construct the prompts used within the study resulted in feedback with a consistent structure that could be programmatically analyzed in almost all cases. This allowed the automation of a substantial part of the evaluation process. Although manual feedback evaluation continued to be necessary for a fraction of them, a way to provide a lower bound on the fraction of erroneous feedbacks that can be obtained automatically was proposed. All these advantages can be leveraged to evaluate different LLMs, different problems on which these models are intended to be applied, different prompts (as long as they provide the feedback with an adequate minimal structure) and even multiple runs for the same prompt (to investigate stochastic effects in the LLM output).

Finally, the structure of the generated feedback is not only useful in evaluation tasks but also in the practical implementation of a hypothetical learning support system, allowing to overcome problems reported in other works, such as handling those situations where the model's feedback contains the code with the solution to the problem (which we do not want to show to the students for pedagogic reasons).

As future work, we find interesting to carry out further research on the development of more advanced ICL-based prompts with improved performance and additional pedagogic properties. In addition to this, we plan to explore the feasibility of implementing feedback quality estimation techniques. These techniques would allow deciding how such feedback or any of its elements should be shown to the students, either by omitting it or warning them in cases where the estimated quality is low.

\section*{Acknowledgments}

This work is supported in part by funds from the Universitat de Barcelona (2023PMD-UB/021 and 2023PMD-UB/026). Authors are also supported by the Artificial Intelligence and Bio-Medical Applications (AIBA) consolidated research group (2021SGR01094) of the Ag\`encia de Gesti\`o d'Ajuts Universitaris i de Recerca (AGAUR).

\bibliographystyle{ACM-Reference-Format}
\bibliography{main}

\end{document}